\documentclass[10pt,final,letterpaper,traditabstract]{swsc_efh}
\pdfoutput=1    
\usepackage[hyphens]{url}
\usepackage{amsmath}
\usepackage{graphicx}
\usepackage{txfonts}
\usepackage[round,sort]{natbib} 
\usepackage[T1]{fontenc}
\usepackage{ifthen}
\usepackage[breaklinks,colorlinks,citecolor=blue,backref=section]{hyperref}
\usepackage[usenames,dvipsnames]{color}
\usepackage[displaymath, mathlines]{lineno} 

\newcommand{\vecr}{\mathbf{r}}
\newcommand{\matR}{\mathbf{R}}
\newcommand{\vectortwo}[2]%
{\begin{pmatrix} #1\\#2\end{pmatrix}}
\newcommand{\vectorthree}[3]%
{\begin{pmatrix} #1\\#2\\#3\end{pmatrix}}
\newcommand{\matrixthree}[9]%
{\begin{pmatrix}#1 & #2 & #3 \\#4 & #5 & #6 \\#7 & #8 & #9\end{pmatrix}}

\newcommand{\degree}{\ensuremath{^\circ}}
\newcommand{\VEV}[1]{\ensuremath{\left\langle{#1}\right\rangle}}
\newcommand{\Ratm}{\ensuremath{R_{\mathrm{atm.}}}}

\definecolor{mycolor}{RGB}{10, 10, 150}

\begin{document}
\volpage{(2017) 7, A24}
\doi{\href{https://doi.org/10.1051/swsc/2017023}{10.1051/swsc/2017023}}
\title{Making 3D Movies of Northern Lights}
\author{Eric Hivon\inst{1}, Jean Mouette\inst{1}, Thierry Legault\inst{2,3}}
\authorrunning{E. Hivon, J. Mouette \& T. Legault}
\offprints{hivon@iap.fr}
\institute{Sorbonne Universit\'{e}s, UPMC Univ.~Paris 6 \& CNRS (UMR7095): Institut d'Astrophysique de Paris, 98bis Bd Arago, F-75014, Paris, France
\and
Astrophotography (\url{http://www.astrophoto.fr}) 
\and
THALES, 19 Avenue Morane Saulnier, F-78140 V\'elizy Villacoublay, France
}
\date{Received 4 June 2017 / Accepted 16 August 2017}

\abstract{We describe the steps necessary to create three-dimensional (3D) movies of 
Northern Lights or Aurorae Borealis out of real-time images taken with two distant high-resolution 
fish-eye cameras. 
Astrometric reconstruction of the visible stars is used to model the optical mapping of each camera 
and correct for it in order to properly align the two sets of images.
Examples of the resulting movies can be seen at \url{http://www.iap.fr/aurora3d}.}


\maketitle

\section{Introduction}
Aurorae borealis have always fascinated humans, who have long tried to report their observations
by the best means available at the time. If the earliest known record of an Aurora was written 
on a clay tablet 567BC in Babylon \citep{Stephenson+2004}, they have since been filmed in color with 
specially designed high sensitivity cameras\footnote{\url{http://auroraalive.com/multimedia/autoformat/get_swf.php?videoSite=aurora&videoFile=aa_aurora_on_tv.swf++&videoTitle=Aurora+on+TV}} in the mid-1980s \citep{Hallinan+1985},
and filmed in 3D for the first time in 2008\footnote{\url{https://www.newscientist.com/article/dn15147-northern-lights-captured-in-3d-for-the-first-time/}}.

In this paper we detail the steps taken to generate 3D movies of Aurorae Borealis shot with 
high resolution wide field cameras, generating movies fit for projection on hemispheric planetariums.
To our knowledge, these 3D movies are the first ones showing Northern Lights ever produced thanks 
to post-shooting astrometric reconstruction.
Since they are produced from real-time movie images instead of the usual time-lapsed sequences, 
they can be used for education and public outreach and give a glimpse 
at the rich and rapidly moving three-dimensional structure of the Auroras.
While they are close to the real-world sensations of Aurora watching in terms of colors, saturation, 
speed of evolution, and sheer size on the sky (if seen in a planetarium), such movies add the extra information
of the stereoscopy, which is inaccessible to a single human observer on the field.
Now that such movies are made available, they will hopefully prompt the interest of scientists studying the Aurorae properties,
who would like to use them, or build upon them, in their work.

The image taking procedure and camera synchronisation is described in Sect.~\ref{sec:observations}.
The image processing steps allowing a good rendition
of the stereoscopy is described in Sect.~\ref{sec:processing}, 
while Sect.~\ref{sec:3Dviewing} shows how such 3D images and movies
can be viewed. Section~\ref{sec:conclusion} is devoted to a conclusion and perspectives.

\section{Observational setup}
\label{sec:observations}
Observations are made from Norway at the approximate latitudes and longitudes of 69.3\degree N and 20.3\degree E, 
with two cameras separated by either 
6 
or 
9
km, depending on the local weather conditions.
Since the Aurorae mostly happen in the upper atmosphere at $\sim100$km above the ground, their parallax is expected to be large enough between the two cameras to construct their 3D rendering via binocular vision.
As much as possible, the cameras point to the zenith, 
and are then rotated around the vertical so that the Northern Star is in the lower part of the image, aligning the long axis of the image with the West-East direction, as shown in Fig.~\ref{fig:camera_referential} for one of the cameras.
Their positions are measured by GPS receivers.
In what follows, the western and eastern cameras are respectively dubbed {\em left} and {\em right}.

\subsection{The cameras}
Each of the two cameras used is a Sony $\alpha$7s, on which is mounted a Canon Fisheye lens of focal length $f$=15mm with 
an aperture of $f$/2.8. In order to further enlarge the field of view, and allow the use 
of a Canon lens on the Sony E-mount camera, it is coupled to a Metabones Speed Booster that reduces the focal length
by a factor of 0.71.
The camera sensor is a full-frame CMOS 4240x2832 pixel RGBG Bayer matrix\footnote{\url{http://www.imaging-resource.com/PRODS/sony-a7s/sony-a7sDAT.HTM}}, with half of the pixels being sensitive to green and the other half spread evenly between red and blue.
ISO is set to 12800, representing a good compromise between noise and sensitivity.
A video signal is produced at 25 or 30 frames per second (fps), in Ultra High Definition format (3840x2160 pixels, ratio 16:9), 
and recorded on an external video recorder, with Apple ProRes HQ codec (approximately 800 Mbits/sec). 
As we will see in Sect.~\ref{sec:processing}, the field of view is 220\degree x110\degree, covering 48\% of the sphere,
with a resolution at the center of the image of 2.94'/pixel.
 
\subsection{Cameras synchronization}
\label{sec:synchronizing}
In order to achieve a good stereoscopy out of the movies produced, we first have to make sure 
that their images are properly synchronized during their processing and visualisation.
Keeping, on the field, the two distant cameras and/or their recorders synchronized thanks to a common absolute
GPS-based time signal would have been ideal, but out of reach of this project. 
Instead,
to achieve a good relative synchronization of the two movies, we first made sure that the internal clocks 
of the two recorders agreed to one second or less before starting every observing nights; we then flashed 
a light in front of the two cameras put side by side and connected to their turned-on recorder 
before setting up the cameras at their respective distant location.
This gives us a good estimate of the offset and possible drift in time between the internal clocks of the
two recorders as they write a time-code in the images metadata.
Finally, in post-production, but before starting the 3D reconstruction pipeline, 
we look in the filmed sequences for the occurence of bright, thin and fast moving Auroras structures,
and compare closely their evolution from frame to frame between the two cameras, 
in order to find the best matching pair of frames.
This provides the residual time offset between the two series of recorded time-stamps, 
and allows us to re-synchronize the two movies at a level compatible with the image rate (25 or 30 fps), 
assuming the relative time drift to be negligible over the duration of a sequence (a few minutes).

\begin{figure}[ht]
\center{
	\includegraphics[angle=0,width=0.45\textwidth,trim={20pt 20 10 40},clip]{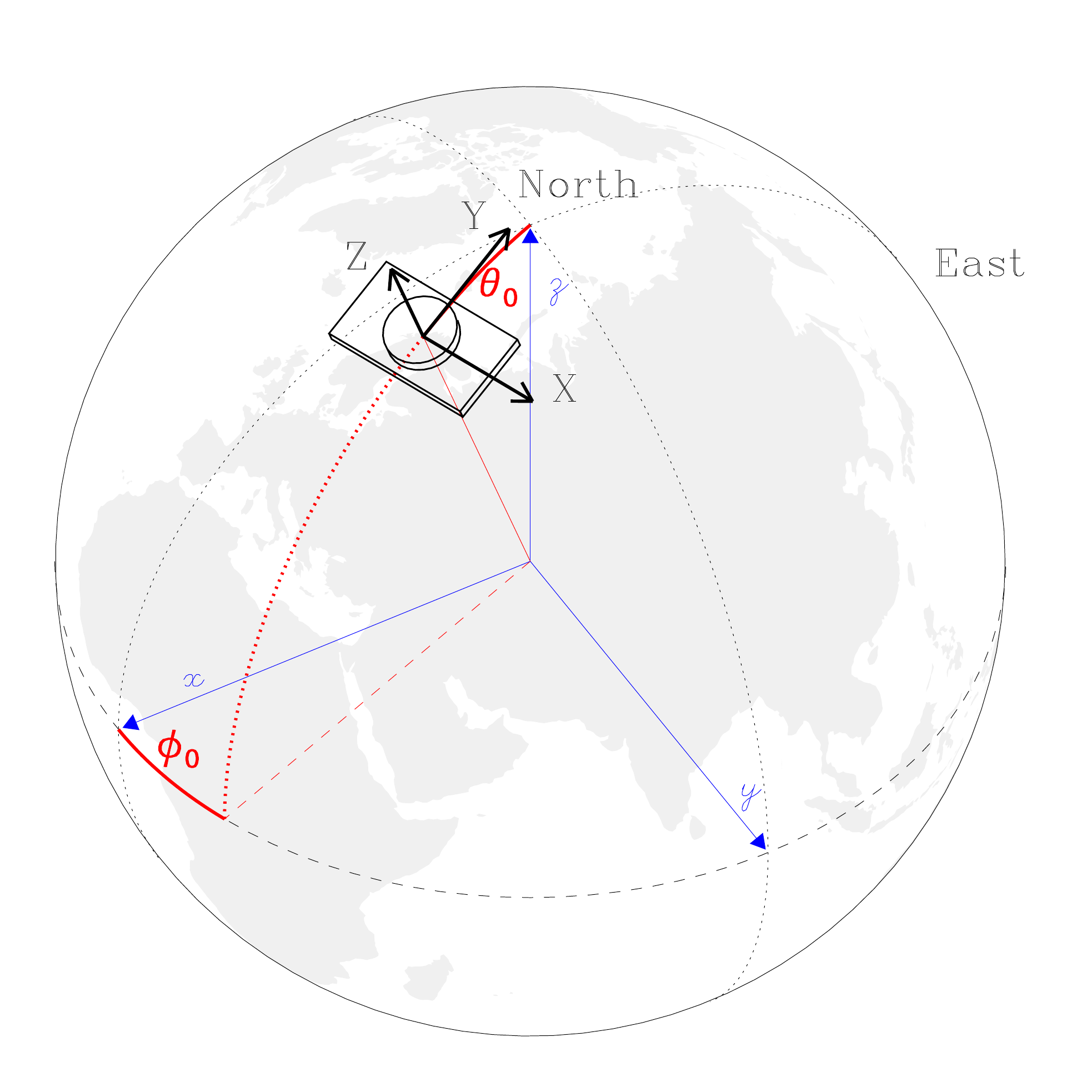}
}
	\caption{\label{fig:camera_referential}
The observation is done from colatitude $\theta_0$ and longitude $\phi_0$, in the Earth-bound ${xyz}$ referential.
A camera-bound referential is defined, with the Z axis along the camera boresight (roughly towards the zenith),
 the Y axis along the shorter dimension of the camera image, approximately pointing North, and
 the X axis along the longer dimension of the camera image, pointing East}
\end{figure}
\section{Image processing}
\label{sec:processing}
Time-lapsed stereoscopic observations of aurorae are described for wide field cameras in \cite{Kataoka+2013}, 
and for fast narrow field cameras in \cite{Kataoka+2016},
following in either cases the procedure described in \cite{Mori+2013}. We are implementing a similar
astrometric method:
since the actual pointing, orientation and relative position of the two cameras are only known
with limited accuracy, and because each camera can distord the image differently,
the positional astrometry of bright stars identified in the images 
is used to determine and characterize
the cameras geometrical settings and optical responses, in order to properly realign the
{\em left} and {\em right} set of images.

\subsection{Finding the stars}
\label{sec:finding_stars}

The 5th edition of the Kitt Peak National Observatory (KNPO) 
catalog\footnote{\url{http://www-kpno.kpno.noao.edu/Info/Caches/Catalogs/BSC5/catalog5.html}}, 
adapted from the 
Yale catalog\footnote{\url{http://tdc-www.harvard.edu/catalogs/bsc5.html}}, 
lists the J2000 equatorial coordinates and magnitude of more than 9000 stars of magnitude $\le 6.5$.
After setting correctly the minus sign on the magnitude of its four brightest stars,
it was used as a catalog of
bright stars potentially seen in the {\em left} and {\em right} images.

Candidate stars in the images are identified as follows. 
Since the Aurorae filmed here are mostly green in color, a red channel frame is expected to be
less exposed to them than the other channels, 
and is used to detect stars. 
We checked that using the green channel, expected to have a better intrisic 
resolution, did not change much the final result, while it increased slightly the risk of false detection.
The chosen image is convolved (via Fast Fourier Transform operations) 
with a difference of Gaussians (DoG) filter, defined in pixel space as
\begin{linenomath*}
\begin{equation}
	F(x,y) = 
		 \frac{\exp{\left(- \frac{x^2+y^2}{2\sigma_1^2}\right)}}{2\pi\sigma_1^2} - 
                 \frac{\exp{\left(- \frac{x^2+y^2}{2\sigma_2^2}\right)}}{2\pi\sigma_2^2}
\end{equation}
\end{linenomath*}
where $x,y$ are integer pixel indices, $\sigma_1 = 1$ and $\sigma_2 = 1.6$,
and one then looks for local maxima in the filtered image.
The value of $\sigma_2/\sigma_1=1.6$ picked here is generally recommended as a good approximation 
of the Mexican hat or Marr wavelet \citep{Marr+1980}, often used to look for point sources in astrophysics 
\citep{Coupinot+1992},
but values as large as  $\sigma_2/\sigma_1 = 3$ gave almost identical results.

For practical reasons, we keep our cameras in focus during the totally of the shooting sequences. 
As a consequence, the Airy diameter 
of the optical system ($\sim 3\mu$m) is smaller than the pixels physical size of the CMOS sensor ($8.40\mu$m), 
and the observed stars generally won't be resolved.
We tried three different techniques to determine the position of a star on the frame:
($a$) that of the locally brightest pixel of the DoG filtered map, 
or the centroid of the filtered map on a ($b$) 3x3 or ($c$) 5x5 patch of pixels
centered on that same brightest pixel. 
For reasons that will detailled in Sect.~\ref{sec:fitting}, the option ($b$) which 
provides sub-pixel accuracy was preferred.
Since it is difficult to estimate exactly the error made on determining the star position, especially since
we start from a (red) image interpolated with proprietary algorithms from the RGBG detector matrix, we will simply assume
the error in position in ($b$) to be the same
as in the discretization scheme ($a$) and determined by the nominal pixel size of the image, 
as described in Sect.~\ref{sec:discretization}. 
Note that over-estimating by a factor 10 the error bar on each star position
won't change the numerical value of any parameter in the multi-dimensional fit
being done, it will only increase by a factor 10 the error bar associated to each parameter.

The $N_b \simeq 40$ brightest sources found by this procedure are compared to the $N_s \simeq 40$ 
brightest stars in the Northern hemisphere, as listed in the input catalog described above.

\begin{figure*}[ht]
\center{
  \includegraphics[angle=0,width=0.65\textwidth,trim={20pt 10 20 10},clip]{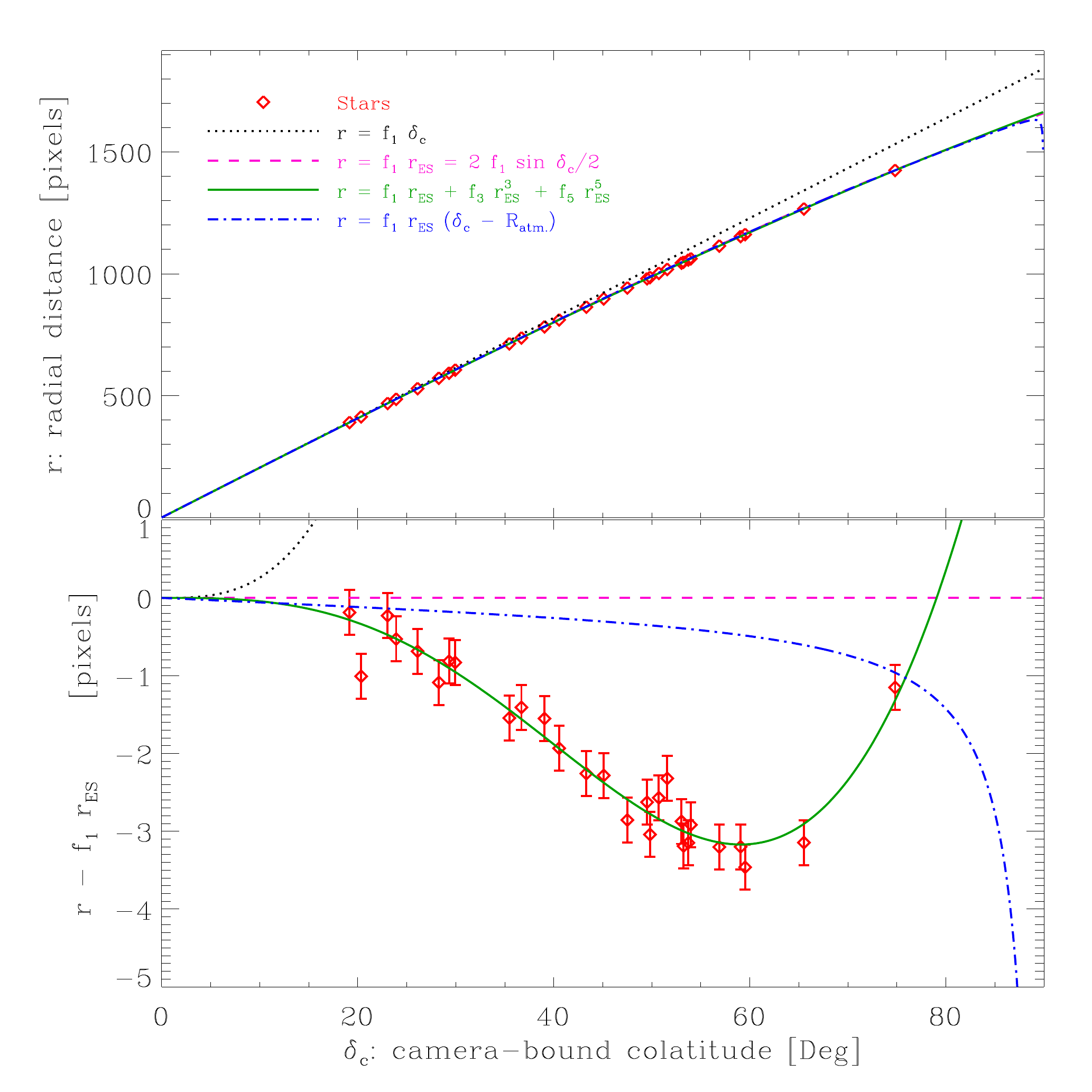}
}
  \caption{\label{fig:radial_model} %
Radial mapping of the camera, relating the camera-bound colatitude $\delta_c$ of a sky object, 
 and its projected radial distance $r$ on the camera image, measured in pixels. 
In the case of stars (red diamonds) the measured radial distance comes with the discretization error bar of size $\pm 1/\sqrt{12} \approx \pm 0.2887$ (see Sect.~\ref{sec:discretization}) more visible on the lower panel.
The black dots show a simple equidistant mapping $r \propto \delta_c$.
The magenta dashes show the equisolid projection of Eq.~(\ref{eq:equisolid}) expected for the fisheye and focal reducer lenses used here;
the green solid line show the best fit distorted radial model of Eq.~(\ref{eq:radial_model});
 while the blue dotted-dashed line illustrate the impact of atmospheric refraction on observations 
with a perfect equisolid optical system (see Sect.~\ref{sec:refraction}). 
The upper panel shows the raw radial distance $r$, 
while the lower panel shows the difference between $r$ and the equisolid model.}
\end{figure*}

\subsection{Modeling the projection}
\subsubsection{3D rotation of the sky}
At a time $t$, a star $(j)$ having the colatitude%
\footnote{In what follows, and unless stated otherwise, the angles will be expressed in degrees; and we will prefer the colatitude $\delta$ measured southward from the North pole to the usual latitude $\delta'$ measured northward from the equator, the two being related by $\delta+\delta'=90\degree$.}
and longitude ($\delta_E^{(j)}, \alpha_E^{(j)}$) in Equatorial coordinates, will have the angular coordinates 
($\delta_c^{(j)}, \alpha_c^{(j)}$) in a camera-bound referential, such as the one illustrated in 
Fig.~\ref{fig:camera_referential}, so that the respective 3D coordinate vectors are 
\begin{linenomath*}
\begin{subequations}
\label{eq:3Drot}
\begin{equation}
	\vecr_i^{(j)} \equiv
	\vectorthree{x_i^{(j)}}{y_i^{(j)}}{z_i^{(j)}} = 
	\vectorthree{\sin\delta_i^{(j)} \cos \alpha_i^{(j)}}{\sin\delta_i^{(j)} \sin \alpha_i^{(j)}}{\cos\delta_i^{(j)}},
\end{equation}
for $i=E$ or $c$.
This change of coordinates is a rotation parameterized by the three (time dependent) Euler angles $\psi, \theta, \varphi$
\begin{equation}
	\vecr_c^{(j)} = \matR(\psi, \theta, \varphi)\ \vecr_E^{(j)},
	\label{eq:rot_coord}
\end{equation}
with
\begin{align}
	&\matR(\psi, \theta, \varphi) \equiv \\
	&\matrixthree	{ \cos\psi}{-\sin\psi}{0}%
			{ \sin\psi}{ \cos\psi}{0}%
			{        0}{        0}{1} \cdot
	\matrixthree	{ \cos\theta}{0}{ \sin\theta}%
			{        0}{1}{        0}
			{-\sin\theta}{0}{ \cos\theta} \cdot
	\matrixthree	{ \cos\varphi}{-\sin\varphi}{0}%
			{ \sin\varphi}{ \cos\varphi}{0}%
			{        0}{        0}{1}.
	\nonumber
	\label{eq:rot_matrix}
\end{align}
\end{subequations}
\end{linenomath*}
The solid-body rotation of the sky described in Eq.~(\ref{eq:3Drot}) is an approximation that ignores subtle distortions like the atmospheric refraction, which will be discussed later on, or the relativistic aberration due to the yearly Earth motion around the Sun. 
The latter is fairly small, with an apparent displacement of the source varying between 0 and $\sim 20.5''$ across the sky, and will affect almost identically the two cameras since they observe the same area of  the sky. 
It should therefore not affect the relative alignment of the images provided by the two cameras.
For a camera located at a time $t$ at the Earth-bound colatitude $\theta_0$ and longitude $\phi_0$, 
pointing exactly to the zenith and rotated by $\psi_0$ around the vertical (with $\psi_0=0$ in the current configuration), 
then, ignoring the Earth nutation,
\begin{linenomath*}
\begin{subequations}
\label{eq:ideal_settings}
\begin{align}
	\psi &=  -90 -\psi_0,\\
	\theta &= -\theta_0,\\
	\varphi &= -\left(\varphi_0 + 15\, S(t) \mod 360\right),
\end{align}
\end{subequations}
\end{linenomath*}
where $S(t)$ is the Greenwich Mean Sidereal Time of observation, expressed in hours, which parameterizes the rotation of the Earth with respect to the distant stars:
\begin{linenomath*}
\begin{equation}
	S(t) = 18.697374558 + 24.06570982441933\,  (J(t)-J_{2000})
\end{equation}
\end{linenomath*}
with $J(t)$ the Julian day of observation and $J_{2000} = 2451545$ is the Julian day starting at 12:00:00 UTC on Jan 1, 2000 \citep{MeeusAstronomical}.
However, as noted previously, the limited accuracy on the actual shooting time, position and orientation of the cameras will impact these parameters. 
The values recovered for $(\psi, \theta, \varphi)$, as described in Sect.~\ref{sec:fitting}, are well within $1\degree$
 of those expected from Eq.~(\ref{eq:ideal_settings}), with the larger discrepancy affecting the harder to set-up azimuthal angle $\psi$.

\begin{table*}[ht]
\center{
\begin{tabular}{ccccccccc}
\hline\hline
Camera & $N_m$ & $f_1$ & $1000 f_3/f_1$  & $1000 f_5/f_1$  & $\Delta_x$ & $\Delta_y$ &  mean discr.  & max discr. \\
\& run &       & $[$mm$]$ &              &                 & $[$pixels$]$  & $[$pixels$]$  &  $[$pixels$]$ & $[$pixels$]$ \\
\hline
L1 & 22 & 9.811 $\pm$ 0.003 & -6.1 $\pm$ 0.5 &  3.8 $\pm$ 0.3 & -17.7 $\pm$   0.5 &   6.0 $\pm$   0.4 & 0.30 & 0.71\\
L2 & 28 & 9.815 $\pm$ 0.003 & -7.0 $\pm$ 0.4 &  4.4 $\pm$ 0.2 & -19.8 $\pm$   0.4 &  -0.1 $\pm$   0.4 & 0.29 & 0.68\\
L3 & 27 & 9.817 $\pm$ 0.003 & -7.4 $\pm$ 0.4 &  4.5 $\pm$ 0.2 & -19.0 $\pm$   0.4 &  -0.0 $\pm$   0.4 & 0.34 & 0.66\\
L4 & 25 & 9.819 $\pm$ 0.003 & -8.2 $\pm$ 0.4 &  5.0 $\pm$ 0.3 & -21.2 $\pm$   0.4 &   0.3 $\pm$   0.4 & 0.31 & 0.58\\
L5 & 22 & 9.818 $\pm$ 0.003 & -6.6 $\pm$ 0.5 &  4.0 $\pm$ 0.3 &  -7.7 $\pm$   0.5 &   2.0 $\pm$   0.3 & 0.41 & 1.07\\
L6 & 23 & 9.859 $\pm$ 0.004 & -5.6 $\pm$ 0.6 &  3.0 $\pm$ 0.4 & -10.2 $\pm$   0.4 &   4.3 $\pm$   0.3 & 0.33 & 0.98\\
\hline
R1 & 23 & 9.890 $\pm$ 0.003 & -5.8 $\pm$ 0.5 &  3.8 $\pm$ 0.3 &  14.0 $\pm$   0.5 &   5.0 $\pm$   0.4 & 0.21 & 0.42\\
R2 & 25 & 9.891 $\pm$ 0.003 & -5.3 $\pm$ 0.5 &  2.9 $\pm$ 0.3 &  13.9 $\pm$   0.4 &   5.3 $\pm$   0.5 & 0.26 & 0.58\\
R3 & 23 & 9.893 $\pm$ 0.003 & -6.7 $\pm$ 0.4 &  4.3 $\pm$ 0.3 &  14.1 $\pm$   0.5 &   4.8 $\pm$   0.5 & 0.22 & 0.44\\
R4 & 23 & 9.885 $\pm$ 0.004 & -3.7 $\pm$ 0.6 &  1.9 $\pm$ 0.5 &  14.2 $\pm$   0.4 &   4.9 $\pm$   0.5 & 0.26 & 0.55\\
R5 & 25 & 9.897 $\pm$ 0.003 & -8.3 $\pm$ 0.4 &  5.6 $\pm$ 0.2 &   8.1 $\pm$   0.5 &   0.4 $\pm$   0.3 & 0.28 & 0.53\\
R6 & 22 & 9.888 $\pm$ 0.004 & -5.3 $\pm$ 0.6 &  3.4 $\pm$ 0.3 &  20.5 $\pm$   0.4 &   8.8 $\pm$   0.3 & 0.33 & 0.85\\
\hline
\end{tabular}
}
\caption{\label{table:red3px}%
Parameters of the camera response, 
for frames taken from six different shooting sequences with the {\em left} 
(rows L1 to L6) and {\em right} (rows R1 to R6) cameras. 
All runs were shot on consecutive nights, except for \#2 to 4 which were produced on the same night, 
and \#6 obtained four months later. Figures \ref{fig:radial_model} and \ref{fig:residual_error} correspond to row R2.
The focal length $f_1$ was converted from pixels to mm assuming a pixel size of 8.40$\mu$m.
The errors quoted on the parameters $f_1,f_3,f_5,\Delta_x, \Delta_y$ 
are those returned by the non-linear fitting procedure, based on the assumed discretization error on star location.
The two rightmost columns show the mean and maximum discrepancy between
the measured and modelled positions of the $N_m$ stars found in the frame.%
}
\end{table*}

\begin{figure*}[ht]
\center{
  \includegraphics[angle=0,width=0.90\textwidth,trim={0pt 5 10 40},clip]{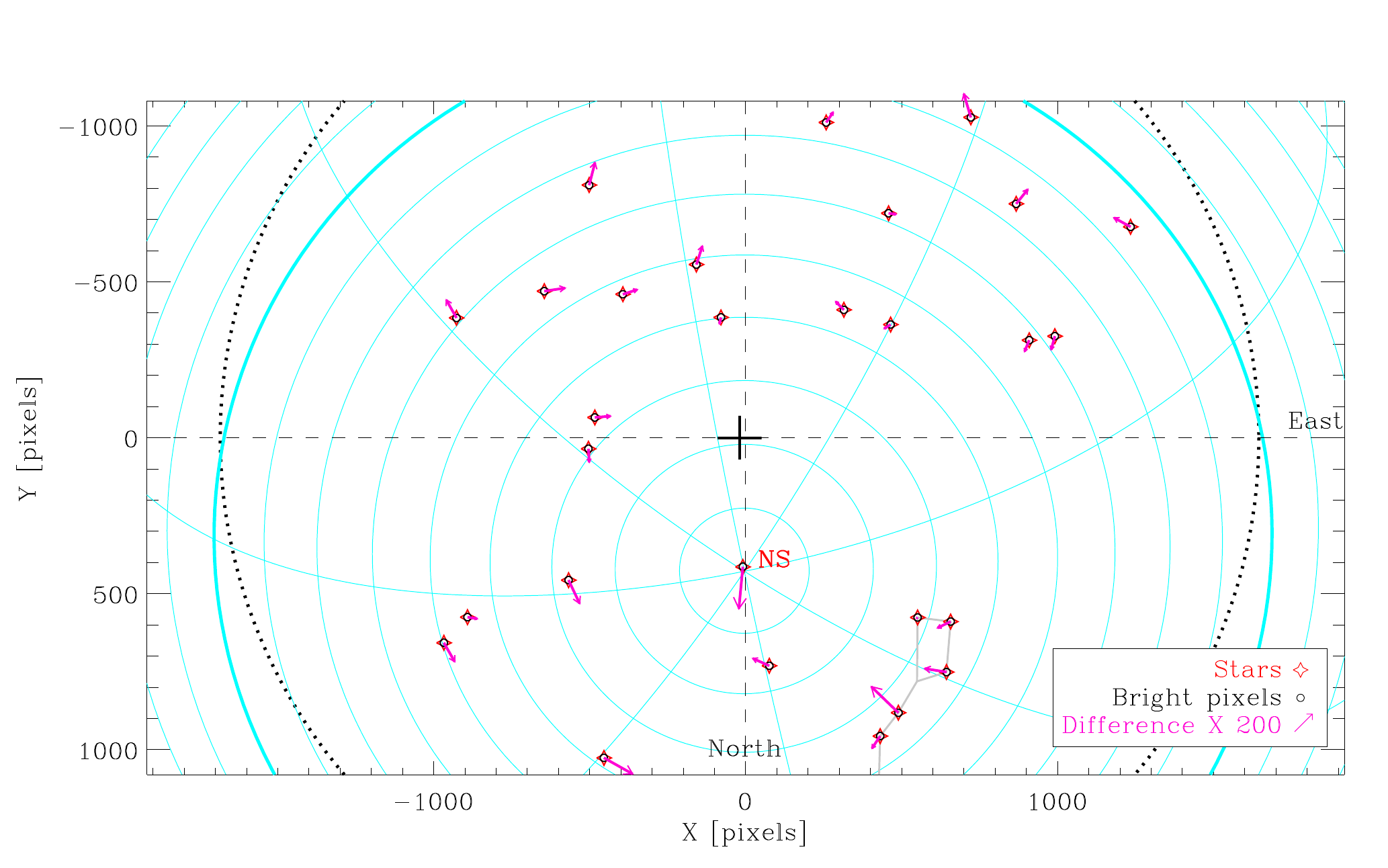}
}
  \caption{\label{fig:residual_error}Comparison of the bright sources location and the position computed with the optical model, for one frame of one of the cameras.
The bright pixels detected in the camera image are shown as black circles, while the expected position of the associated stars, computed by Eq.~(\ref{eq:projection}) with the eight camera parameters obtained in Sect.~\ref{sec:fitting}, are shown as red stars. 
The residual discrepancies between the two, {\em multiplied by 200}, are shown as magenta arrows. 
The thick black dots outline the approximate landscape horizon on the image;
the black cross close to the image center shows the optics focal point;
and the cyan lines represent the Equatorial graticule, with a spacing of $10\degree$ in (Equatorial) colatitude and $45\degree$ in longitude. 
The Northern Star is marked NS, while the ``dipper'' part of {\em Ursa Major}, in the North-East quadrant, is outlined in grey.}
\end{figure*}

\subsubsection{Radial distortion of the optics}
\label{sec:radial_distortion}
A source located at an angular distance $\delta_c$ of the camera bore-sight will appear on the
imaging CMOS sensor at a distance $r$ of the focal point, 
as illustrated in Fig.~\ref{fig:radial_model} and we model this distance as
\begin{linenomath*}
\begin{subequations}
\label{eq:radial_distortion}
\begin{equation}
	r^{(j)} = f_1 r_{ES}(\delta_c^{(j)}) + f_a r_{ES}^a(\delta_c^{(j)}) + f_b r_{ES}^b(\delta_c^{(j)}),
	\label{eq:radial_model}
\end{equation}
where
\begin{equation}
	r_{ES}(\delta_c) \equiv 2 \sin(\delta_c/2),
	\label{eq:equisolid}
\end{equation}
\end{subequations}
\end{linenomath*}
is the Equisolid radius {\em ideally} expected for the fisheye and reducer lenses being used, 
$f_1$ is the combined focal length
of the fisheye and focal reducer, and $f_a$ and $f_b$ model any non-linear departure 
from the expected projection, for the integer numbers $a$ and $b$.
We found the combination $(a,b)=(3,5)$ to be slightly better than $(2,3)$, by requiring smaller amplitude of the corrections
ie, $f_3/f_1 \simeq -0.008$ and $f_5/f_1 \simeq 0.004$, instead of $f_2/f_1 \simeq -0.02$ and $f_3/f_1 \simeq 0.01$, in the range
probed by the stars ($\delta_c < 90\degree$ and $r_{ES}< 1.41$), and by returning slightly smaller residual in the comparison
of the model with the observations.

Since the optics is pointing toward the zenith, these non-linear 
terms can result from both radial non-idealities of the optics and, as suggested by \citet{Mori+2013}, 
the atmospheric refraction of the incoming light, which increases with the camera-bound colatitude of the source.
However, as shown in Sect.~\ref{sec:refraction} and in Fig.~\ref{fig:radial_model}, atmospheric refraction is too small in amplitude to explain the distortions seen, since
we would then get $f_3/f_1 = 2 f_5/f_1 \simeq - 8.24\,10^{-5}$.

\subsubsection{2D offset of the camera sensor}
Finally, the geometric center of the camera sensor, identified as $(x,y)=(0,0)$ 
may not match exactly the optics focal point, and we allow for the shifts
$\Delta_x$, $\Delta_y$, so that a source $j$ will appear on the sensor at the location 
\begin{linenomath*}
\begin{subequations}
\label{eq:CMOS_shifts}
\begin{align}
	x^{(j)} &= r^{(j)} \cos \alpha_c^{(j)} + \Delta_x,\\
	y^{(j)} &= r^{(j)} \sin \alpha_c^{(j)} + \Delta_y,
\end{align}
\end{subequations}
\end{linenomath*}
measured in units of pixels, from the nominal CMOS sensor center. 
As listed in Table~\ref{table:red3px}, 
we find these shifts to be respectively $|\Delta_x|\sim 20$ and $|\Delta_y|\sim 5 $ pixels, 
depending on the camera considered, and the conditions of observation.

\subsection{Fitting the parameters}
\label{sec:fitting}
Combining Eqs.~(\ref{eq:3Drot}), (\ref{eq:radial_distortion}) and (\ref{eq:CMOS_shifts}), the position of
a star on the camera sensor is related to its Equatorial coordinates via
\begin{linenomath*}
\begin{align}
	\vectortwo{x^{(j)}}{y^{(j)}} = D\left(\psi, \theta, \varphi, f_1, f_a, f_b, \Delta_x, \Delta_y; 
	\vectortwo{\delta_E^{(j)}}{\alpha_E^{(j)}} \right),
\label{eq:projection}
\end{align}
\end{linenomath*}
where the eight parameters ($\psi, \theta, \varphi, f_1, f_a, f_b, \Delta_x, \Delta_y$) are assumed unrelated
between the two cameras. 

In order to determine these parameters for a camera, we begin by setting the angles ($\psi, \theta, \varphi$) to the values given in Eq.~(\ref{eq:ideal_settings}), and letting all the other parameters to 0, except for $f_1$ which is set to the nominal focal length of the optical system.
Applying Eq.~(\ref{eq:projection}) to the $N_s$ brightest stars of the catalog, we find that for each of a handful of them ($N_m$), 
the computed location is within a few pixels of a bright source identified on the image. Assuming the star and 
the close bright pixel to be the same object, we then look for the set of parameters that minimize the $2N_m$ 
discrepancies in $(x,y)$ coordinates. 
To do this non-linear fitting, we tried and compared two different IDL implementations 
of the Levenberg-Marquardt algorithm \citep{marquardt1963, numrec}: the {\tt curvefit}\footnote{\url{https://harrisgeospatial.com/docs/CURVEFIT.html}} routine included in the standard IDL 
library, and its drop-in replacement {\tt mpcurvefit}\footnote{\url{http://purl.com/net/mpfit}}
by \citet{markwardt2009} 
based on the {\tt MINPACK} algorithm of \citet{More1978}.
We found the best fit parameter values to be nearly identical, but only the latter routine 
returned meaningful error bars on those parameters.
Injecting these new parameters in Eq.~(\ref{eq:projection}), the number of coincidences $N_m$ is increased, 
and we again look for a new estimate of the parameters minimizing the $2N_m$ discrepancies. 
The process is repeated a few times, rapidly converging toward a stable number of matches 
($22 \le N_m \le 28$ depending on the image being treated), and providing a stable set of fitted parameters.
 
Figure~\ref{fig:residual_error} compares the measured position of the bright sources and the computed position 
of their matching stars in the best fit model of the camera response, for one specific image. 
Their discrepancies, shown as magenta arrows 
(after multiplication by 200) seem fairly randomly distributed and do not exhibit any clear trend.
The residual distances have a mean value of $0.29$ (in pixel units), with a worst case of $0.68$. 
The mean residual found is compatible with the error created by 
assigning an integer value to the star coordinate in the image, which is $\simeq 0.3826$, as shown in Sect.~\ref{sec:discretization}.

As shown in Table~\ref{table:red3px}, the same reconstruction procedure was applied, for each of the two cameras, 
on images extracted from six different shooting sequences, 
filmed on four different nights spread over a four month period. 
The optics related parameters ($f_1, f_3, f_5, \Delta_x, \Delta_y$) and the level of residual discrepancies in position,
 were found to be quite stable for each camera, 
with the largest relative changes affecting $\Delta_x, \Delta_y$ when the lenses were unmounted and remounted between observations
performed on different nights.
The relatively large changes of $f_3$ and $f_5$ between runs may be due to 
a partial (anti-)correlation between these two parameters, which also shows
on the fact that the combination $f_3+f_5$ varies less between runs than either $f_3$ or $f_5$. 
However, this degeneracy 
does not hamper the modelling of Eq.~(\ref{eq:projection}) as long as $f_3$ and $f_5$ are treated together, 
and not considered separately. This suggests however that any more sophisticated modelling of the optical response,
and in particular of the radial distortion,
would require either the use of a basis of orthogonal polynomials, or a physically motivated set of parameters.

In each case, the mean residual error in position is compatible or below what is expected from quantization error 
of the star coordinates (0.3826 pixels), and we note that the {\em right} camera 
seems to perform a bit better than the {\em left} one
on this respect. This may be due to the slightly lesser sharpness of the images produced by the {\em right} camera, 
maybe due to its optics,
which by bluring lightly the stars makes the sub-pixel determination of their position easier.

As mentioned in Sect.~\ref{sec:finding_stars}, different options were considered to determine the stars position
on the camera frame. Using the centroid on a 5x5 patch (so-called option ($c$)) 
instead of 3x3 used here made no significant difference on the
parameter values and did not improve the residual discrepancies compared to those listed in Table~\ref{table:red3px}.
As expected, using instead a discrete pixel location (option ($a$)) 
lead to different parameter values, with a shift compatible with the quoted error bars,
and larger mean and worst case discrepancies (by $\sim$30\% and $\sim$20\% respectively). 
However, for the 12 images tested, the sky to pixel 
mappings of Eq.~(\ref{eq:projection}) resulting from options ($a$) and ($b$) respectively 
differed by quite less than one pixel in the region of interest, 
ie, everywhere above the horizon, making the resulting
processed images and movies of Auroras nearly indistinguishable.

The 8-parameter mapping model considered here therefore seems to generate smaller reconstruction errors
than the 6-parameter model considered in \cite{Mori+2013}, which ignores the non-linear radial distortions ($f_a$ and $f_b$).
On the other hand, in the framework of meteors and transient objects detection,
\cite{Borovicka+1995} proposed a model containing 13 (12 independent) parameters 
for the modelling of all sky cameras equiped with photographic plates featuring a measurement uncertainty of $\sim$1 arcmin.
With hour-long exposures and $\sim$100 stars for calibration, they reached a residual modelling error of $\sim$1 arcmin too.
We did not investigate how such a model would have performed in our case.
Even though adding more parameters is not problematic in itself for the non-linear fitting procedure, it 
may lead to further degeneracies between parameters, 
unless they are carefuly designed to be orthogonal, especially if the number of constraints (here, the identified stars)
is not large enough to correctly probe each of them.
Moreover, inspection of Figs.~\ref{fig:radial_model} and \ref{fig:residual_error} does not suggest the need for more degrees of freedom.

\begin{figure}[ht]
\center{\vbox{
  \includegraphics[angle=0,width=0.49\textwidth,trim={0pt 0 0 0},clip]{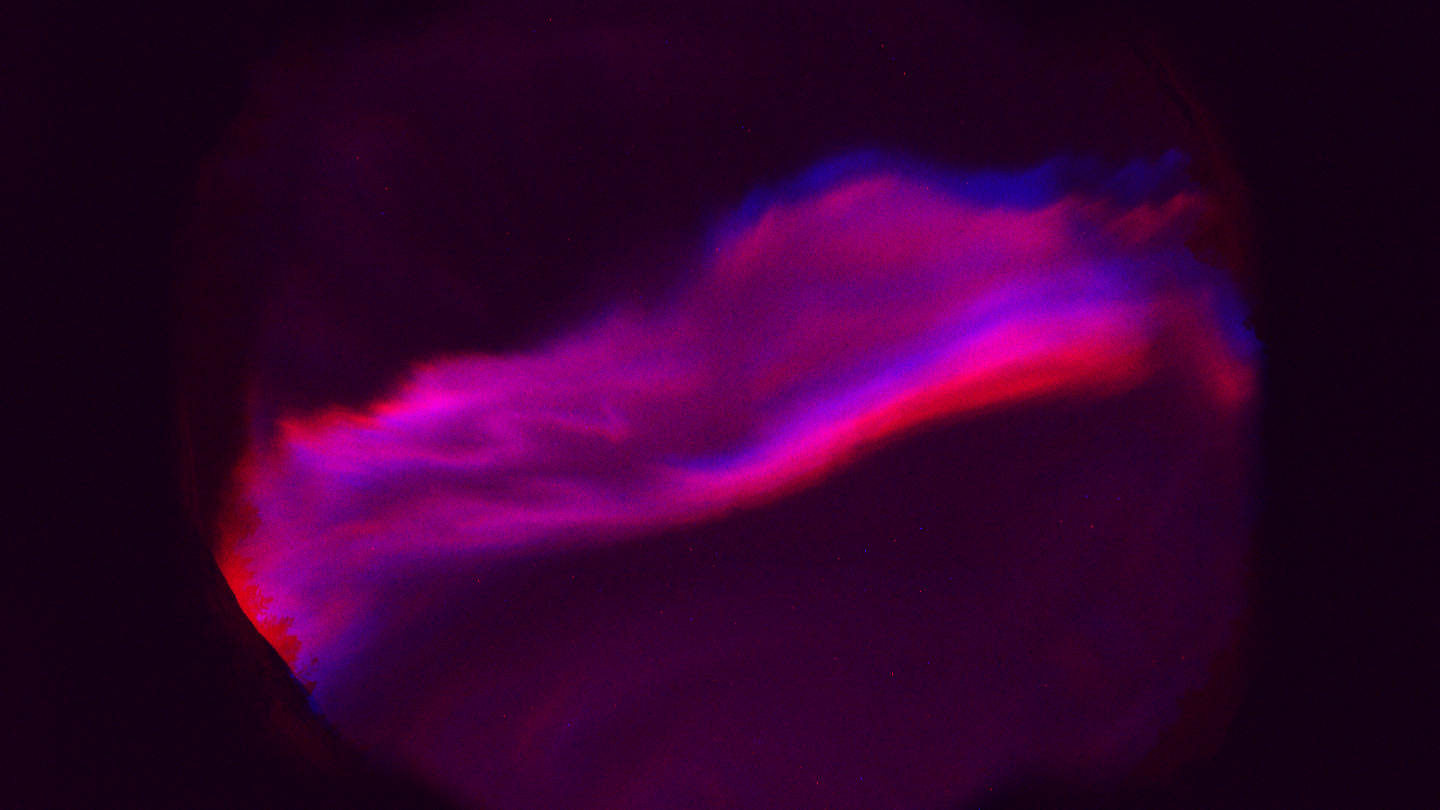}\\
  \includegraphics[angle=0,width=0.49\textwidth,trim={0pt 0 0 0},clip]{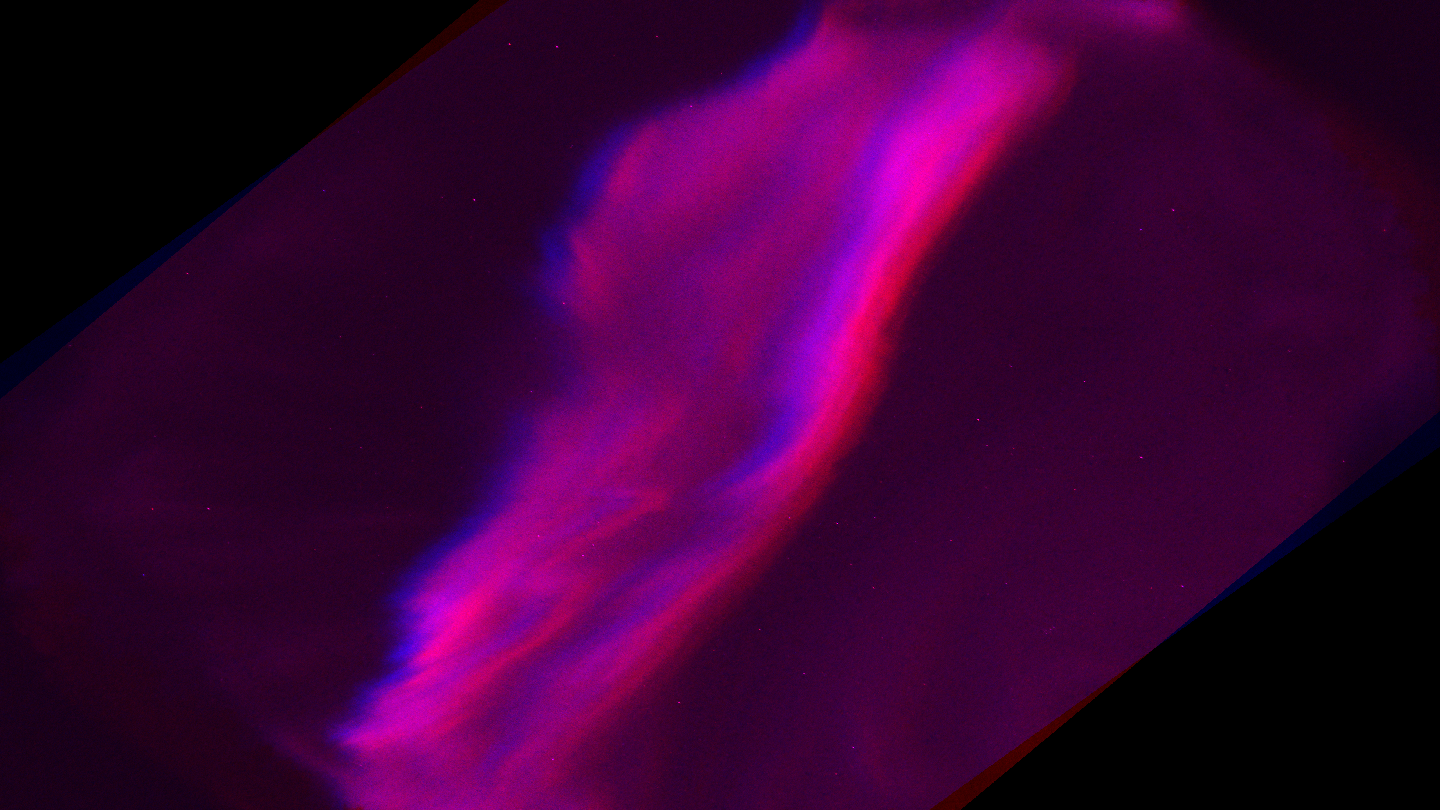}
}}
	\caption{\label{fig:anaglyph}Anaglyph rendition of an Aurora snapshot, 
 before (upper panel) and after (lower panel) alignment of the images produced by the {\em left} and {\em right} cameras}
\end{figure}
\subsection{Aligning the images}
Once the parameters are determined for each of the {\em left} and {\em right} images, 
with respect to the absolute Equatorial coordinates,
one can map the two images (or set of images) 
onto a common projection of our choice, so that the stars match exactly for the {\em left} and {\em right} ``eyes'',
while the structures of the aurorae will be offset according to their parallax.
We chose for instance, for flat screen viewing, and for Figs.~\ref{fig:anaglyph} and \ref{fig:freeviewing},
to deproject the {\em right} image by inverting Eq.~(\ref{eq:projection}) 
for the {\em right} camera parameters, 
and projected it again by applying that same equation with the {\em left} camera parameters. Other choices are possible,
such as mappinng the two images onto a common equidistant projection, 
as required by the Dome Master format\footnote{\url{www.mayaskies.net/production_tools/articles/Introducing_the_dome_master.pdf}} used in planetariums.

Apart from the angle $\varphi$, which includes the apparent motion of the sky due to Earth rotation,
the parameters are not expected to vary during the typical duration of a filmed sequence (2 to 3 minutes).
We chose to determine these two sets of parameters on the first image of the {\em left} and {\em right} movies respectively
and apply the mappings based on them, one frame at a time, to the whole span of each of the two movies,
meaning that the stars will appear to slowly rotate on the sky, 
as they would for a human observer with a very wide field of view.

An extra twist is that the two cameras do not sit on the same latitude, even though the long axis of
the images they generate are along the East-West axis. It is therefore necessary to rotate the two sets of images 
by (almost) the same angle so that the direction of parallax matches the horizontal 
axis of the screen on which they will be projected.

Figure \ref{fig:anaglyph} illustrates the impact of the images alignment and rotation.

\section{3D viewing}
\label{sec:3Dviewing}
\subsection{Stereoscopic techniques}
The 3D images generated above have been tested with different stereoscopic techniques,
which are listed below in the order of increasing hardware requirement and quality of
the rendition.

\subsubsection{Free-viewing and cross-eye}
\begin{figure*}[ht]
\center{\hbox{
  \includegraphics[angle=0,width=0.49\textwidth]{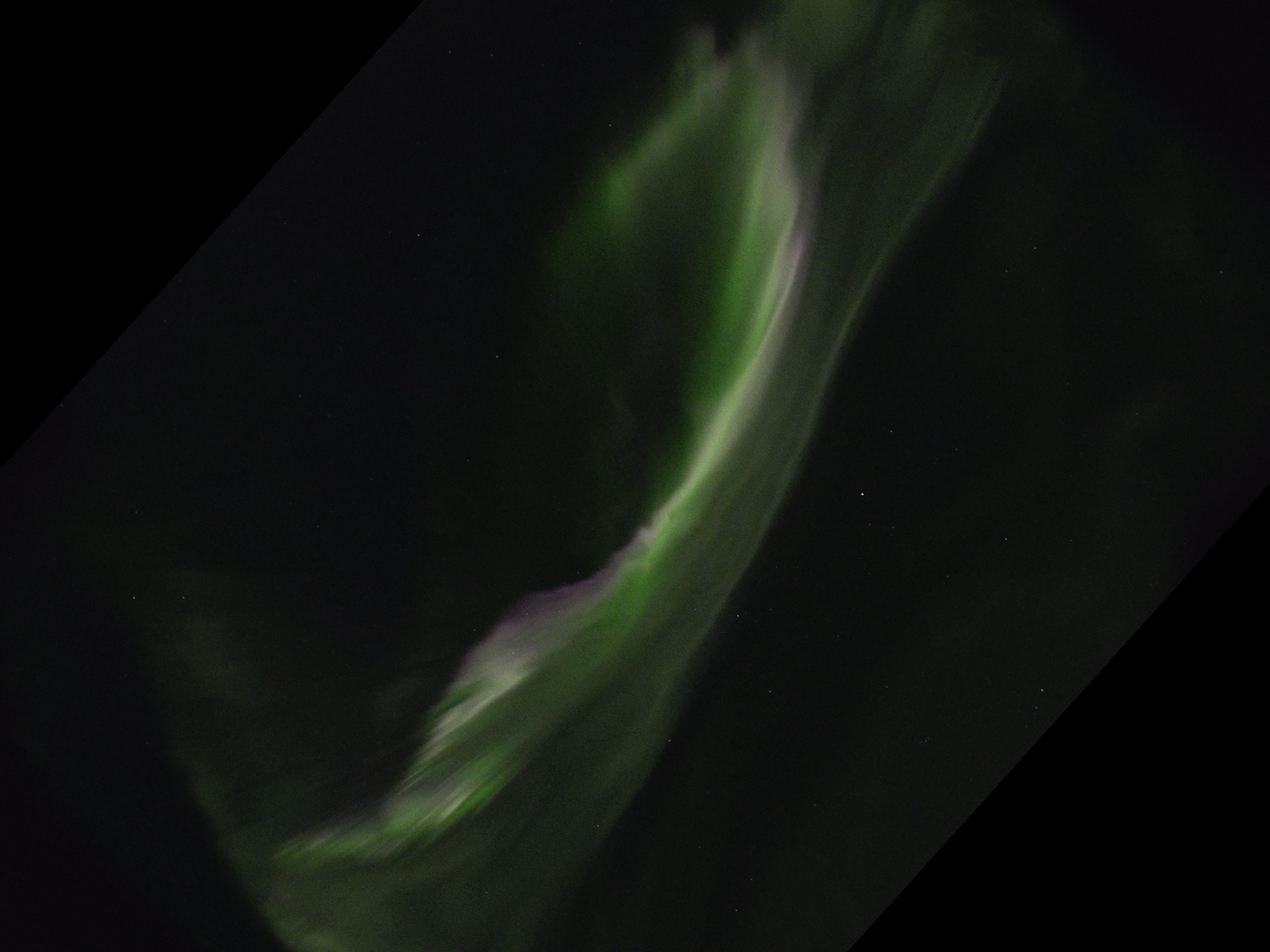}
  \includegraphics[angle=0,width=0.49\textwidth]{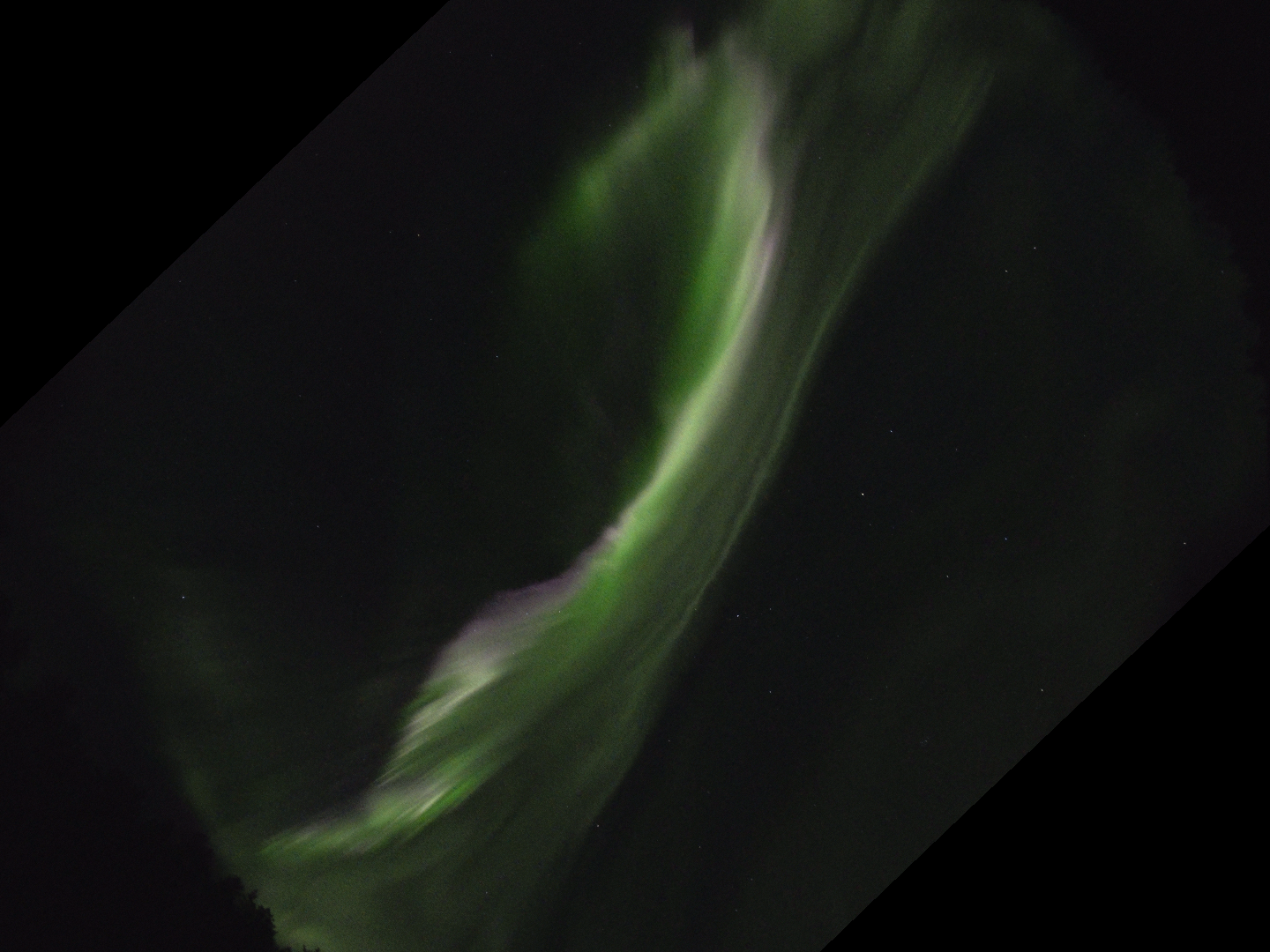}
}}
	\caption{\label{fig:freeviewing}{\em Left} and {\em right} images of an Aurora, 
cropped to a 4:3 aspect ratio, for 3D free-viewing, as described in Sect.~\ref{sec:3Dviewing}.
The cross-eye technique, where the {\em left} and {\em right} images are swapped, can be tried by turning the image up-side down}
\end{figure*}

The {\em left} and {\em right} images are shown side by side, either at their respective location 
and watching far beyond the screen, in the so-called free-viewing technique, as shown in Fig.~\ref{fig:freeviewing},
or after swaping their position (ie {\em left} image on the right side and vice-versa), 
and crossing the eyes, in the aptly named cross-eye technique.
These techniques, requiring no material, are by far the cheapest and most accessible ones, 
but take some time to master, and are limited to very small images.

\subsubsection{Anaglyphs}
Since the Auroras are mostly green in color, a single composite image of the realigned observations
is made in which the red channel is the green component of the {\em left} image, and the blue channel is 
made out of the {\em right} image.
This is looked at with anaglyph glasses 
in the which the left glass is red, and the right one is blue.
This technique, available on screen and in print, as illustrated in Fig.~\ref{fig:anaglyph}, 
is extremely cheap, 
with the drawback of loosing color information and reducing luminosity.

\subsubsection{Passively polarized screen and glasses}
On an ad hoc screen, the odd lines, polarized vertically, 
show every second line of the {\em left} image while the even lines, polarized horizontally, do the same for the {\em right} image. 
The left and right lenses of a pair of glasses are polarized vertically and horizontally respectively.
More recent models of screens and glasses use circular polarization instead.
This technique, used on some 3D TV screens, is affordable and preserves the colors, but has a resolution cut in half.

\subsubsection{Time multiplexing and active glasses}
The {\em left} and {\em right} images are shown alternatively on a LCD screen polarized 
at 45\degree{} and supporting high frequency refreshment rate, 
and the left and right glasses become alternatively 
opaque and transparent to that polarization, thanks to rotating liquid crystals, 
with a shutter system operating in synch with the screen.
This technique, preferred for computer video games, requires specific glasses 
and proprietary software and video card.
Like all techniques based on linear polarization, it is limited to small or medium size screens.

\subsubsection{Other techniques}
Other techniques that we were not able to test with our Aurorae images, 
but are extremely suitable for very large screens,
include time multiplexing of circularly polarized images, shown in alternance,
with circularly polarized passive glasses, 
very widespread in 3D movie theaters, or
wavelength multiplexing, where the {\em left} and {\em right} images use differents red, green and blue wavelength bands,
to which the left and right dichroic lenses are respectively transparent, and which is mostly used in planetariums.\\
Yet another pathway is the use of virtual reality glasses, allowing the immersion of the viewer in a 
outdoor scene illuminated by 3D Northern Lights. However, for maximum efficiency this requires
a heavier observational set-up providing two different views of, at least, the whole hemispheric sky.
Set-up that we are only starting to implement.

\subsection{Depth of images}
In all the techniques listed above, the stars, which after the image processing of Sect.~\ref{sec:processing} 
match exactly on the {\em left} and {\em right} images, will appear 
to be on the plan of the paper or of the screen, with the Aurora floating in front of them.
Depending on the technique used and the size of the image, and in order to improve the feeling 
of immersion, it may be necessary to move the whole 3D image 
closer to or further from the viewer. This can be achieved by shifting, for instance, 
the {\em left} image toward the right or the left respectively.

\section{Conclusion}
\label{sec:conclusion}
In this paper we present the image processing pipeline implemented to 
produce real time 3D movies of Aurorae Borealis out of images produced by two distant high resolution fish-eye 
cameras.
A model of the camera optical response is proposed, 
and for each camera, its eight numerical parameters are constrained using
the positional astrometric information of the two dozen bright stars it observed. Each frame of
the filmed rushes can be then corrected from this response in order to properly superpose the images, and
produce 3D movies with a good stereoscopy.
Samples of minutes-long movies produced by this pipeline are available for download at \url{http://www.iap.fr/aurora3d}, 
and can be watched on monitors or projection screens adapted for 3D vision, 
as described in Sect.~\ref{sec:3Dviewing}, and probably in virtual reality headsets.
The IDL source code of the pipeline, and its fish-eye projection model,
will be made available at the same location, 
after it has been properly cleaned-up and documented.

We are now applying the same techniques to images and films produced with two pairs of camera, 
separated by $\sim$6km as previously, and
the cameras in each pair set up so that the whole hemispheric sky is observed at once by each pair.
After careful stitching of the images, this provides higher quality 3D movies suitable 
for hemispherical screens, such as planetariums, or for virtual reality glasses,
at the cost of heavier observational setup and logistics.

We hope that the existing movies, and the forthcoming ones, will be of interest to the general
public raptured by the spectacle of Northern Lights, as well as to scientists studying the phenomenon.

\begin{acknowledgements}

We thank Fran\c{c}ois S\`evre and St\'ephane Colombi for insightful discussions, 
Fran\c{c}ois Bouchet for funding this article publication charges,
Inge Petter Amlien and Elisabeth Nyg{\aa}rd for their hospitality and friendship in the far North,
as well as Urpo and Tarja Taskinen for their cozy welcome on the Arctic polar circle.

The authors and the editor thank two anonymous referees for their very constructive comments and suggestions, 
which helped improving this paper and for their assistance in evaluating it.
\end{acknowledgements}

\appendix
\section{Impact of atmospheric refraction}
\label{sec:refraction}
The atmospheric refraction makes astronomical objects appear higher above the horizon 
(ie, at smaller {\em local} colatitude, measured from the zenith)
than they actually are, with an offset increasing with the colatitude.
In the simplest model, 
in which the atmosphere is approximated as a uniform and flat slab of gas,
a source of actual colatitude $\delta$ will appear at a colatitude $\delta_a$ shifted by
\begin{linenomath*}
\begin{equation}
	\Ratm =  \delta - \delta_a = R_{45} \tan( \delta_a ) \approx R_{45} \tan( \delta ),
	\label{eq:atmos_refraction}
\end{equation}
\end{linenomath*}
and \citet{Comstock1890} proposed that the amplitude of refraction, 
parameterized by its value at $45\degree$, $R_{45}$, expressed in arcsec, depends on the observational conditions via
\begin{linenomath*}
\begin{equation}
	R_{45} =  \frac{21.5\, P}{273 + T},
\end{equation}
\end{linenomath*}
with $P$ and $T$ the atmospheric pressure and temperature, in mmHg and $\degree C$ respectively.
One then notes that, for small values of $\Ratm$, the apparent equisolid radial distance is
\begin{linenomath*}
\begin{subequations}
\begin{align}
	r_{ES}(\delta - \Ratm) &\simeq r_{ES} - \Ratm \cos \delta/2, \\
			& =r_{ES} -  R_{45} \frac{r_{ES}}{1-\frac{r_{ES}^2}{4-r_{ES}^2}}, \\
			& = (1-R_{45})r_{ES} - R_{45} \sum_{p\ge1} \frac{r_{ES}^{2p+1}}{2^{p+1}}.
\end{align}
\end{subequations}
\end{linenomath*}
At sea level, for high pressure ($P=800$mmHg $=107$kPa) and low temperature ($T=-20\degree C$), $R_{45} = 68'' = 3.3\, 10^{-4}$rad,
and the apparent location of a source seen through a fisheye of focal $f$ with perfect equisolid projection would be
\begin{linenomath*}
\begin{equation}
	r(\delta)/f = 0.99967 r_{ES} - 8.24\,10^{-5} r_{ES}^3 - 4.12\,10^{-5} r_{ES}^5 + \ldots
\end{equation}
\end{linenomath*}
Of course, Eq.~(\ref{eq:atmos_refraction}) is very crude model of atmospheric refraction, 
but for sources $10\degree$ or more above the horizon, it only differs by a few arcsec at most from
more accurate and complex models.
Moreover, this model most likely over-estimates the refraction close to the horizon since
it diverges at $\delta =90\degree$, leading us to consider that the actual refraction will be smaller than the one estimated above.

\section{Discretization error}
\label{sec:discretization}
Since the projection of the optics is close to equisolid, as illustrated in Fig.~\ref{fig:radial_model},
it preserves the areas and therefore the surface densities.
If the distribution of stars is assumed uniform on the sky, it will also be uniform on the image.
When assigning a star position $(x,y)$ to the closest pixel of integer indices $(i,j)$
the error made is $(dx,dy) = (x-i,y-j)$, where $dx$ and $dy$ both have a probability 
density of 1 on the interval $]-1/2,1/2]$.
Therefore, the means and variances of these errors are $\VEV{dx}=\VEV{dy} = 0$ and $\VEV{(dx)^2}=\VEV{(dy)^2} = 1/12$.
Defining the actual and approximate radial distances as
$r \equiv \left(x^2+y^2\right)^{1/2}$ and $r' \equiv \left(i^2+j^2\right)^{1/2}$ respectively, then their difference
\begin{linenomath*}
\begin{subequations}
\begin{align}
	dr &= r-r',\\
	&\simeq\frac{ (dx)^2 + (dy)^2 }{2 r'} + \frac{i dx + j dy}{r'} \quad \mathrm{for}\ r' \gg 1,
\end{align}
\end{subequations}
\end{linenomath*}
 has for mean value $\VEV{dr} = 1/(12 r')$
and for standard deviation 
\begin{linenomath*}
\begin{align}
	\sigma_{dr} = 1/\sqrt{12} \simeq 0.28867.
	\label{eq:sigma_dr}
\end{align}
\end{linenomath*}

One also notes that $s \equiv (dx)^2$ has the density $\rho_{s} = 1/\sqrt{s}$ for $0\le s \le 1/4$, 
and therefore 
\begin{linenomath*}
\begin{equation}
	e^2 \equiv (dx)^2 + (dy)^2
\end{equation}
\end{linenomath*}
 has the density
\begin{linenomath*}
\begin{subequations}
\begin{align}
	\rho_{e^2} &= \pi             \quad \quad \quad \quad \quad {\mathrm{for}} \  0 \le e^2 \le 1/4, \\
		   &= 2 \arcsin (\frac{1}{2e^2}-1)\quad {\mathrm{for}} \ 1/4 < e^2 \le 1/2,\\
		   &=0 \quad \quad \quad \quad \quad {\mathrm{otherwise.}}
\end{align}
\end{subequations}
\end{linenomath*}
It implies 
\begin{linenomath*}
\begin{equation}
	\VEV{e} = \left(\sqrt{2} + \ln\left(1+\sqrt{2}\right)\right)/6 \simeq 0.3826,
	\label{eq:mean_error}
\end{equation}
\end{linenomath*}
and 
\begin{linenomath*}
\begin{equation}
	\sigma_e = \left( 1/6 - \VEV{e}^2 \right)^{1/2} \simeq 0.14242.
\end{equation}
\end{linenomath*}

\bibliography{noli3dv5_bib}

\end{document}